# Perspective: Ultrafast magnetism and THz spintronics

Jakob Walowski, Markus Münzenberg

*Institut für Physik, Ernst-Moritz-Arndt-Universität Greifswald; 17489 Greifswald, Germany*

**This year the discovery of femtosecond demagnetization by laser pulses is 20 years old. For the first time this milestone work by Bigot and coworkers gave insight in a very direct way into the time scales of microscopic interactions that connect the spin and electron system. While intense discussions in the field were fueled by the complexity of the processes in the past, it now became evident that it is a puzzle of many different parts. Rather than giving an overview that has been presented in previous reviews on ultrafast processes in ferromagnets, this perspective will show that with our current depth of knowledge the first real applications are on their way: THz spintronics and all-optical spin manipulation are becoming more and more feasible. The aim of this perspective is to point out where we can connect the different puzzle pieces of understanding gathered over 20 years to develop novel applications. based on many observations in a large number of experiments. Differences in the theoretical models arise from the localized and delocalized nature of ferromagnetism. Transport effects are intrinsically non-local in spintronic devices and at interfaces. We review the need for multiscale modeling to address processes starting from electronic excitation of the spin system on the picometer length scale and sub-femtosecond time scale, to spin wave generation, and towards the modeling of ultrafast phase transitions that altogether determine the response time of the ferromagnetic system. Today, our current understanding gives rise to the first real applications of ultrafast spin physics for ultrafast magnetism control: THz spintronic devices. This makes the field of ultrafast spin-dynamics an emerging topic open for many researchers right now.**

The rise of ultrafast magnetism began with the observation of the ultrafast switching in nickel observed by Bigot and coworkers in Strasbourg 20 years ago[1]. It was a major breakthrough and the experiment challenged the fundamental understanding of magnetism at that time. So far, nanosecond experiments on gadolinium, the magnetic system was, in some ways, seen to be independent of the electronic system. Therefore, the fastest way to convert heat to the spin system was thought to be the wiggling of the lattice. Heat transfer channels were thought to be via spin-orbit coupling, which manifests as magnetic anisotropy that couples the spin direction to the lattice-orbitals. Magnetic anisotropy energy is, however, rather small in the meV range corresponding to a time scale larger than 10 ps. Therefore, the demagnetization time below 100 fs was very surprising. The short time is a direct evidence for the strong connection of spins and electrons that react much faster, on the 10 fs time scale. Like an ultramicroscope for microscopic processes, it tells us something about the initial steps of spin-orbit scattering in the electronic bands and the physics of spin-flip processes, about Stoner excitations and exchange scattering, the building blocks for spin-wave dynamics at THz frequencies and spin transport on femtosecond time scales. However, the observed processes are still hidden in a few parameters (e.g. demagnetization time of the magneto-optical Kerr response). This made the research field both exciting and complex early on, however it also shows the demand for the development of further experimental techniques providing deeper insight and clever experiments to disentangle these processes. This introduction highlights the complexity that makes the dynamics of the system multi-faceted. One simple theoretical approach cannot give a complete description of the diversity of processes happening, and more exciting discoveries in ultrafast magnetism are on the horizon. In this perspective, we make some simple initial considerations on a textbook introductory-like level to get the reader into the stage of current discussion based on some simple blueprints. From here we are ready to enter a discussion on new developments of THz spintronic applications and give a perspective on experiments, theoretical models and future developments of ultrafast magnetism. For extensive reviews please refer to Kirilyuk et al.[2] and Bigot et al.[3].

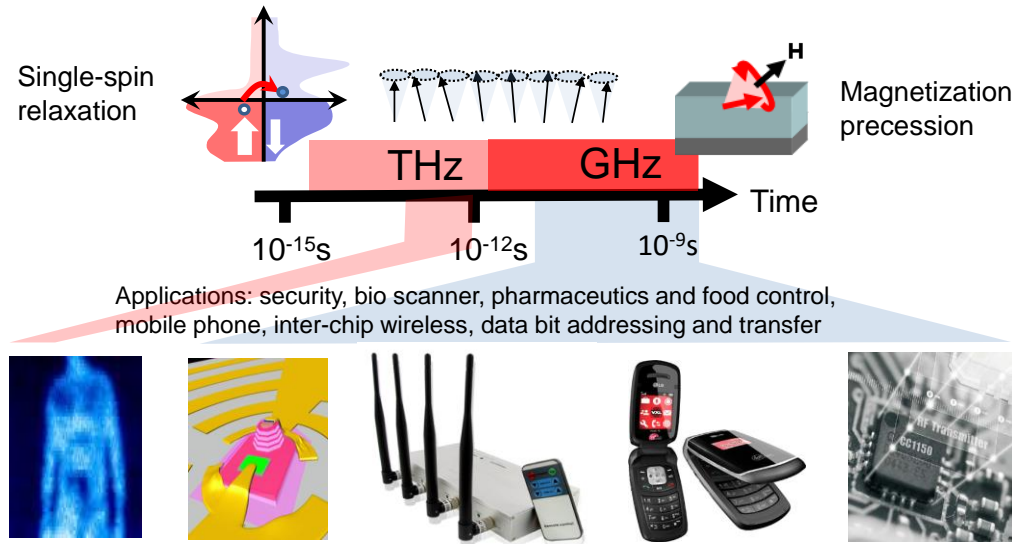

**Fig. 1: Time scales of spin-dynamics.** From single spin-flips and spin waves on THz frequencies to GHz magnetization precession (top). These are connected to possible future spintronic applications and devices from GHz to THz frequency generators (bottom). Reprinted from Miao et al.[52].

**I. Introduction: Current understanding of ultrafast processes**

What is this insight alike we get through experiments on ultrafast timescales? Many observations in a large number of experiments over the last years [4,5,6,7,8,9,10,11,12,13,14] reveal us different insights, and since typical electronic excitations are found on femtosecond time scales, fundamental discoveries of the solid state can be made in the femtosecond region. Femtosecond laser experiments can be compared with particle accelerator experiments in nuclear physics aiming to break the ground state into fundamental excitations. In solid state physics we observe the fundamental mechanism of scattering and energy dissipation. To understand one of the challenges in the field of ultrafast magnetism in the past, one has to keep in mind that magnetism is a manifestation of the Coulomb interaction and the Pauli principle. Describing spin-spin interaction by one parameter called the exchange interaction, $J_{ex}$, was a genius construction by Heisenberg[15] and Dirac[16]. This gives instructive insights into magnetic ordering of spin systems. However, one has to keep in mind that this ignores the underlying detailed electronic structure. On the other hand, early methods developed by Stoner[17] allowed calculating first spin-split densities of states including electronic features of the bands. However, these were not calculated relativistically and the method neglects the spin-orbit interactions, which are very important in ultrafast magnetism.

We need the aforementioned fundamental approaches, which lead to a separation of electrons and spins, to describe the complex interactions and spin-dynamics in the ferromagnet if they are applicable. For example, the mapping of a complex electronic system onto the spin properties in the form of a Heisenberg exchange, $J_{ex}$, leads to an atomistic spin model[18]. The spins at each atom site, interacting with their neighbors can be described using the Landau-Lifshitz-Gilbert equation of motion, derived from the basic quantum mechanical Zeeman term including some viscous energy dissipation of the ensemble[19]. This is a powerful, predictive method in nanomagnetism. Nevertheless, the artificial separation is challenging our understanding when we have to think about the interaction of the excited electron system, and then mapping these dynamics onto the spin system in a second step. Similar to electron transport in a metal in presence of spin-orbit interaction, as for example in case of the spin-Hall effect[20], one can map the crystal state onto the spin-quantum number. To what extend is this useful? In the solid state crystal, due to the spin-orbit interaction, the spin is not a good quantum number and the related state not an eigenstate. A projection of the proper electron state onto the spin quantum number is therefore questionable[21]; the electron state is rather a mixture of a spin-up and spin-down state[22,23,24,25]. Similar effects are found in the presence of spin-fluctuations. However, this spin-mixing has important implications. Even for a propagating Bloch wave without any scattering, the spin's orientation, and thus its momentum, is not conserved. Bloch states driven by light, electric fields or scattering will propagate the electron's spin and orbital state in time into states with a different mixture of spin-up and spin-down. Many of the misunderstandings at the onset of this field are related to the fact that the spin is not a good quantum number in a crystal. A separation into spin-up and spin-down states and arguments on momentum conservation have to be taken with care. Spin-orbit effects are central in describing the evolution of spin states in driven systems and, if we control them, for excitation and detection of spins currents. Spin-orbit interaction is the essential ingredient of understanding ultrafast magnetization dynamics.

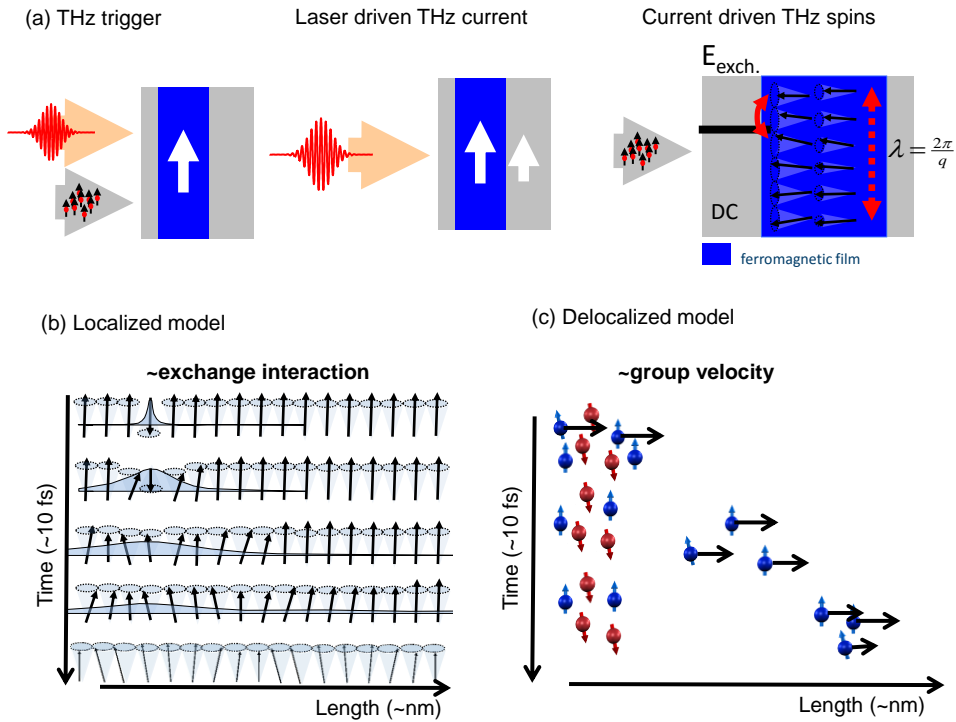

**Fig. 2: Trigger for THz dynamics: spin currents and spin waves.** (a) The THz time scale can be imprinted by a femtosecond laser pulse triggering picosecond current burst (laser driven THz current), or by a THz spin-wave mode (current driven THz spin waves). (b) Local spin excitation decaying into a spin-wave shown schematically in a time-space diagram. (c) Excited spins in the delocalized model have different speeds or decay constants, which results in a spin-polarized current. Time scales are dominated by the group velocity or diffusion speed that differ for spin-up and spin-down electrons, shown schematically in a time-space diagram. Similar time scales are found in both of these simple pictures.

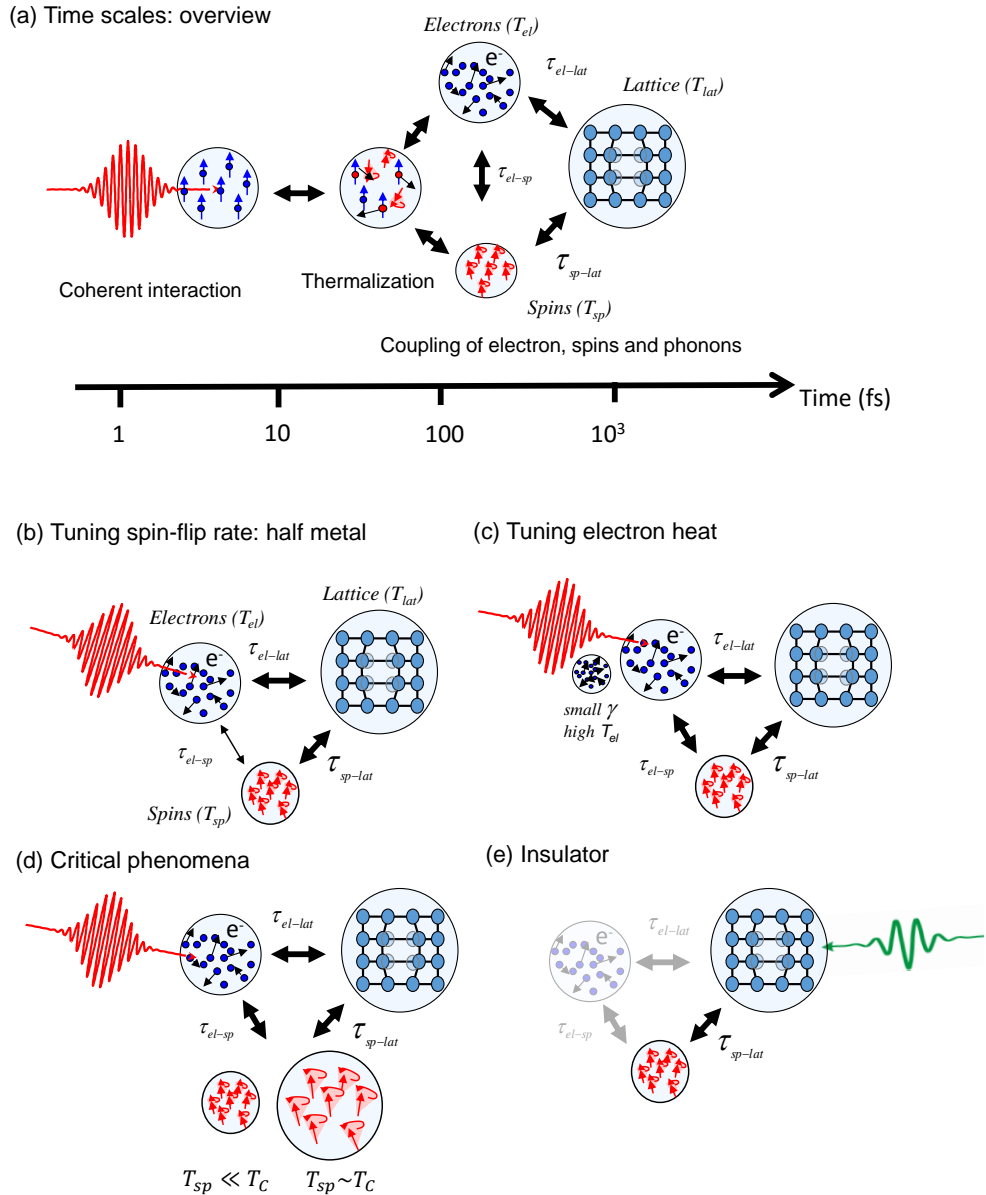

**Fig. 3: Schematics of the time scales of laser driven interaction versus a time ray from 1 fs to 1 ps.** (a) Non-thermalized distributions are depicted. Thermalized distributions of electrons and spin ensembles are assumed after >50-100 fs (adapted from Bigot et al.[150]). Modified 3T model for a simplified schematic description of generic effects. (b) Decoupling of spin and electron system in low damping systems (Heusler alloys, Half metals). (c) Variation of the electron specific heat (large increase of electron temperatures for small Sommerfeld parameter $\gamma$) (d) Observation of critical phenomena for high electron temperatures reaching the Curie temperature ($T_C$) (strong increase of spin specific heat). The ratio of the specific heats is depicted schematically by the respective area. In (e) for magnetic insulators (e.g. yttrium iron garnet (YIG)) pumping phonons by THz radiation is shown, resulting also in a fast demagnetization.

## A. Rate equations versus ultrafast spin transport

The separation of electrons, spins and phonons can give some initial helpful insights into ultrafast magnetism, explaining the general dynamics observed. If we neglect the underlying details of the interactions, general trends can be derived. A rate model of coupled equations describes the time scales of energy transfer between three subsystems for electrons, spins and phonons that can be compared to the time-resolved reflectivity dynamics and time-resolved Kerr data. This allows the interaction and equilibration rates in between electrons, spins and phonons to be extracted. What are the preconditions for this approximation? As Born and Oppenheimer[26] pointed out, a separation of atom dynamics of the nucleus and the surrounding electrons can be made since the electrons will adiabatically follow any slow changes of the lattice. Typical dynamics of the phonon system is at around ~1ps. Similarly, a separation of the charge and the spin degree of freedom is reasonable since slower time scales are observed for the dynamics of the spins, ~100 fs, than that for the electron scattering, ~10 fs. Time scales can be quite different for each of these processes. Electron scattering is energy dependent typically in the range of 1-50 fs, spin wave dynamics is found from THz to GHz depending on the magnon wavelength and typical phonon modes are found to reach the THz range as well, depicted in Figures 1 and 2. Thus, a use of this oversimplified approach seems to be hopelessly inadequate. One reason that this approach nevertheless describes the general appearance of the dynamics is that all these microscopic excitations, the degrees of freedom of the system in the ensemble, average.

The microscopic processes in their ensemble can be described by macroscopic variables and their temperatures: specific heat ($C_{el} \sim \gamma T_{el}$, $C_{sp}$, $C_{ph}$) and scattering rates ($\tau_{el-sp}$, $\tau_{el-ph}$, $\tau_{sp-ph}$)[1]. Using the three coupled rate equations and a delta- or Gaussian excitation, still analytical solution can be found and used to extract the equilibration times in between the subsystems[27]. In Figure 3 (b)-(e), we show four exemplary scenarios to demonstrate the general effects of their interaction. In Figure 3 (b), if the electron-spin interaction is set to zero or small value, then the spin system's temperature will not be dominated by the coupling to the electron system, but by the spin-lattice interaction, which is much slower. In Figure 3 (c) for a small electron specific heat $c_{el} \sim \gamma T_{el}$, with the Sommerfeld coefficient $\gamma$, the electron temperatures will be very high at the initial stage when all energy is deposited by the optical excitation herein. In Figure 3 (d), the spin specific heat at around the phase transition, the Curie Temperature $T_C$, increases in the ferromagnet. This increase is connected with critical phenomena at the phase transition and results because of the strong increase in the specific heat and a delayed increase of the spin-system temperature. In Figure 3 (e), in an insulator, the electron system cannot be excited directly. Currently, heating by THz radiation coupling resonantly to specific phonon modes is investigated by different groups and reveals also a fast response of the spin system[28]. This coupling will result in an increased phonon temperature and equilibration with the spin-system is determined by the spin-lattice coupling. The drawback is that all these rate models always rely on the separation of electrons, spins, lattice and thermalized distributions of the excitation spectra. In addition, in reality parameters can have quite some temperature dependence, energy dependence or be different for the non-thermal electronic distribution, as schematically illustrated in Figure 3 (a). However, we benefit from the fact that the spin-mixing is low, only a few percent. Although, small portions for example in Ni of 4% of the states can have even 20% spin-mixing in the Brillouin zone[29]. This simplification still allows separating spin-up and minority spin-down channel for the greater part of electrons. The spin-mixing is then present in the interaction channel between electron and spin systems, described as a second order process.

In spin transport, the separation of transport into spin-up and spin-down channels is known as Mott's two current model[30], with one current flowing for spin-up separated from the current flowing for spin down electrons in separated bands with individual conductivity. The same microscopic scattering rates appear[31], as described in the previous section, and are well known for the description of giant-magnetoresistance effects[32],[33]. This strong connection to spin transport should also be visible in other effects. Indeed M. Battiato et al.[34] suggested in 2010 that, in addition to this local equilibration of the subsystems, transport should be connected arising from nonlocal effects: spatial gradients in the laser excitation will result in transport effects to equilibrate the energy distribution laterally after the laser pulse hits the sample and excites it locally, as shown in Figure 2 (c). First experimental evidence had been published by Melnikov et al.[35]. This ballistic or diffusive spin and electron transport can be thought of as an ultrafast spin-dependent Seebeck effect[36],[37],[38]. The denomination of a ballistic or diffusive transport regime depends on length scales and spin scattering rates. In the regime of strong scattering, the transport is diffusive[39]. Between diffusive and ballistic transport, the system is described as superdiffusive[40], that means that few particles can propagate ballistically without scattering for longer distances, while the majority is scattered. The phenomena of these rare events are called Lévy flights[41], leading to a different power law in the diffusion equation. Generally, the power law for a mean electron displacement is $<\Delta r(t)^2> \sim D\, t^\alpha$, where D is the diffusion coefficient and t is the elapsed time. In superdiffusion, the few electrons that can travel undisturbed over long distances lead to $\alpha > 1$. As a consequence, ultrashort laser pulses can be applied to trigger ballistic or diffusive spin-and electron transport with picosecond rise times on nanometer length scales. This opens up completely new possibilities for ultrafast magnetism and spin electronics, merging into the novel field of THz spintronics.

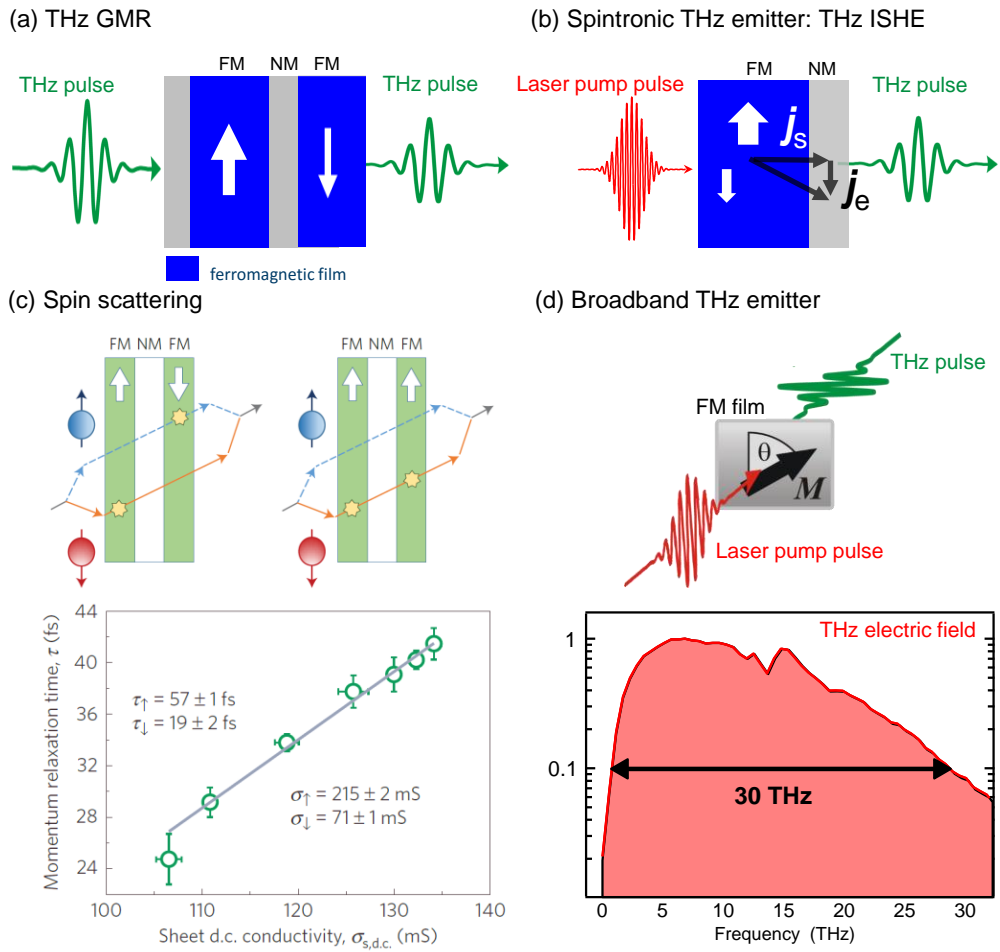

**Fig. 4: Prospective THz spin transport and spin-orbit-based devices.** (a) Schematics of the THz giant magneto resistance (GMR) and (b) of the THz inverse spin Hall effect (ISHE) in ferromagnet (FM) nonmagnetic metal (NM) layers. (c) Accessing fundamental processes of spin transport (Drude relaxation) by the THz GMR and (d) THz spintronic application in broadband THz emitter based on the ISHE, from [53,63,64].

In the case of the superdiffusive spin currents, the role spin polarized currents is of importance. A thermally driven spin polarized current originates from different Seebeck coefficients in the two spin channels. The effect had been first observed by Slachter et al.[42] and called spin-dependent Seebeck effect. In contrast the spin Seebeck effect is a net spin pumping current over the ferromagnet/metal interface induced by a non-equilibrium magnon distribution most prominent if the ferromagnet is an insulator as yttrium iron garnet (YIG)[36,43]. THz magnons in a fs-pump-probe scenario could also pump pure spin currents at interfaces, as discussed in a s-d model by Tventen et al.[44].

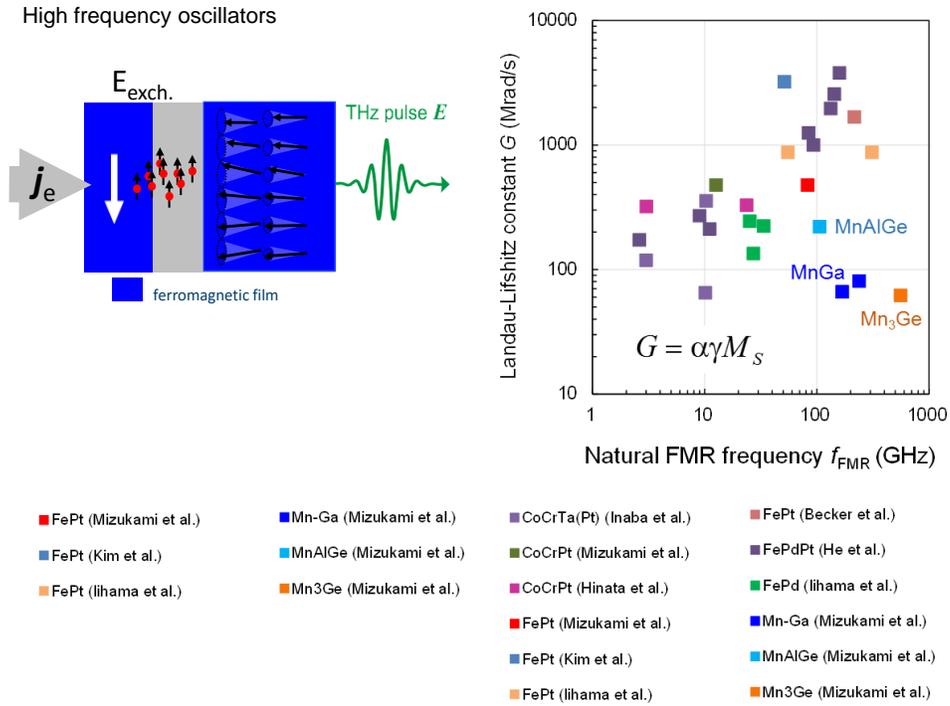

**Fig. 5: THz oscillators.** Low damping, high frequency oscillators is of utmost importance for THz applications using spin waves. Recent optimization of materials is shown, from [75,76].

Currently, applications from magnetic tunnel junctions to recent spin-Hall devices, spin-orbit torque and heat related spin-Seebeck effects, are seeding the field of spintronics and the emerging field of orbitronics[45]. We now understand the processes on ultrafast timescales well enough to develop novel devices exploiting spin-dependent and spin-orbit effects. All spintronics and emerging orbitronics devices can find their counterpart on ultrafast time scales. The future of field of THz spintronics promises very rich in investigations to come.

## II. Novel applications in the THz range

### A. THz spintronics

Why might THz spintronics be interesting for computing? Although current semiconductor transistor developments with 12 nm gate-pitch face serious leakage currents and power consumption is increasing. To a certain extent, on-chip power management allows power consumption to be balanced, especially for mobile devices[46]. Spintronic devices are an option to reduce power consumption. Spin-based RAMs (Spin Torque Random Access Memory (ST-MRAM) from Everspin[47] currently serve as embedded memory or special high reliable automotive solutions. These can be integrated as the top layer into 40 nm and 28 nm CMOS processors (for perpendicular-MTJ Spin Torque eMRAM (embedded))[48]. Potentially, spintronic computational devices have the advantage of non-volatility, low writing currents, and at the same time, high reliability. This means that future breakthroughs will probably go through a change of paradigm, like three-dimensional chip structures or using plasmons or magnons for computing[49]. Another way would be to speed up computing. Computational power is given by the number of operations made per unit time and unit area. In the last years, Moore's law was fulfilled for the footprint area of the transistor. The number of transistors per area is still increasing and the current 22 nm lithography node that uses partly three-dimensional gates, as for example Intel's 3D tri-gate transistor, is already moving towards the third dimension[50]. However, frequency clocking of the devices has remained at the same level since the year 2000 at a few GHz. By using THz spintronics, one could open new avenues in computational speeds by combining ultrafast optics and photonics with spintronics. A closer synchronization of processing and memory clock could be achieved by exploiting a THz spintronic memory and processor. THz spintronic technologies are not only interesting for ultrafast-computing, THz is on its way to becoming a sophisticated-spectroscopy tool[51], as ultrafast lasers become more available, and THz technologies for security are almost installed by now at every airport in North America. THz spintronics is enabling easy-to-use THz devices that are close to being implemented into applications today.

THz spintronics opens a new paradigm with current ultrafast technologies. A schematic of THz spintronic devices is shown in Figure 2 (a). One needs on one hand ultrafast control. This can be achieved through ultrafast light, heat, magnetic field or electric field pulse to trigger the picosecond process, and on the other hand one needs an ultrafast readout. If a laser pulse is used, as shown in the middle of Figure 2 (a), this excites an electron bunch. Non-equilibrium electrons will drive current though the device and ferromagnetic layers, leading to picosecond spin or charge current bunches. On the right, a process is shown where a current can drive coherent spin excitations at fixed frequency. Spin-wave resonators in the GHz range are standard meanwhile. However, the field of spin-wave resonators in the THz range is opening up. THz-spintronic modulators for currents can be based on magnetic tunnel junctions with low damped high-frequency spin wave modes. For THz spintronic applications, as suggested by Miao et al. in 2011 [52], both processes can emit THz radiation. They widen the spintronic frequency range into the THz range for applications depicted in Figure 1.

To transfer the concepts of spintronics to the THz frequency range, we have to prove that the basic concepts of spintronics are still valid at THz time scales. A special test case is the giant magneto resistance (GMR), for which the Nobel prize was awarded to Albert Fert and Peter Grünberg [32,33]. Indeed, it can be demonstrated that standard spinelectronic phenomena work at THz frequencies, (Figure 4 (a)). The GMR effect still operates at THz frequencies: Mott's two current model, with different relaxation channels on different spin channels, has been proven to be functional and different Drude relaxation times can be extracted for spin-up and spin-down electron transport[53] (Figure 4 (c)). The static GMR of 23% is comparable to the THz GMR of 25%. In addition, individual microscopic spin scattering channels can be determined. Also the spin-transfer torque effect is operative on THz frequencies[54]. Magnetic RAMs currently are operated by magnetic tunnel junction based bits. It is essential to manipulate magnetization in such bits in a controlled way by the application of a spin torque. In magnetic tunnel junctions, the bias voltage delivers an additional control parameter: tuning the potential difference, and thereby the transport through the barrier. This allows a selection of states available in tunneling. Under different bias-voltages, the spin-dynamics have been studied using ultrafast lasers in search of a modification of the ultrafast demagnetization[55]. Spin-transfer torque driven by extreme temperature gradients seem feasible [56,57] as indicated by research published so far [58,59].

New concepts in spintronic devices focus on a more compact version of the magnetic device structures. This is possible by 'bending' electrons and exploiting the third dimension. Rather than have the writing and the readout process both in the same element, the magnetic tunnel junction, in a linearly constructed, sequential device, spin-orbit based effects allow to separate the writing and the readout processes. In these so-called spin-orbitronic devices, the spin-current for switching the memory layer is produced by the spin-orbit torque effect. At the metal/ferromagnet layer of the base contact, a current through the material drives a charge to spin-current conversion[60]. The optimization of these devices and the understanding of the spin-orbit driven spin-currents for the static case is just the beginning. Two different processes, an intrinsic conversion and spin-current injection in two steps[61] or Rashba-like interface effect[62] seem both to be relevant for the spin-orbit torque generation. How about the THz timescales? In Figure 4 (b) and (d), a double layer device is shown that works on this principle. We look here at the effect of a spin-current coming from the ferromagnet though the interface that is then converted into a charge current. Because of the different nature of the bands in the ferromagnet owing different relaxation times and different Fermi velocities for the two spins, the laser pulse drives a spin current in a ferromagnet. In a second step, the spin current is injected into the metal layer. In the presence of spin-orbit interaction, the inverse spin-Hall effect 'bends' the electrons depending on their spin orientation. This transfers the spin current transmitted through the interface into a charge current, perpendicular to the spin-current and spin polarization direction. A charge current bunch generated in this way can be directly measured, since the picosecond current bursts radiate like antennas and emit THz electromagnetic waves from the sample [63,64]. The electromagnetic wave emitted can be sampled. It is a measure for the current flow in the sample, an ultrafast ammeter. At the same time, these devices can be optimized for their THz emission amplitude and band width. Because of the large currents generated in the highly non-equilibrium situation, and the fast time scales of the current burst resulting from the small length scales, this gives spintronic THz emitters unique properties. The power of the THz emission is comparable to standard THz emitters (GaP, ZnTe) and only a factor of five smaller compared to low-temperature GaAs based emitters[65,63], which are photoconductive switches that generate picosecond current pulses in defect-rich semiconductor materials. Those are currently used for airport security scans and commercial spectrometers. However, the bandwidth of a spintronic THz emitter is much larger, 1-20 THz compared to the latter with 1-3 THz only. This opens up new possibilities for the improvements of THz spectrometers for bio- and medical applications and for spectroscopic fingerprints in search of explosives. Also, the polarization of the THz pulse is easily modified by the magnetization direction of the emitter structure, allowing easy means to control linearly and circularly polarized THz electromagnetic waves, and will allow the development of powerful THz near field sources[66,67]. Recently, control of the THz emitted electromagnetic waves by the laser pulse's polarization state had been demonstrated[68]. Both gives new and easy means to control THz radiation emission.

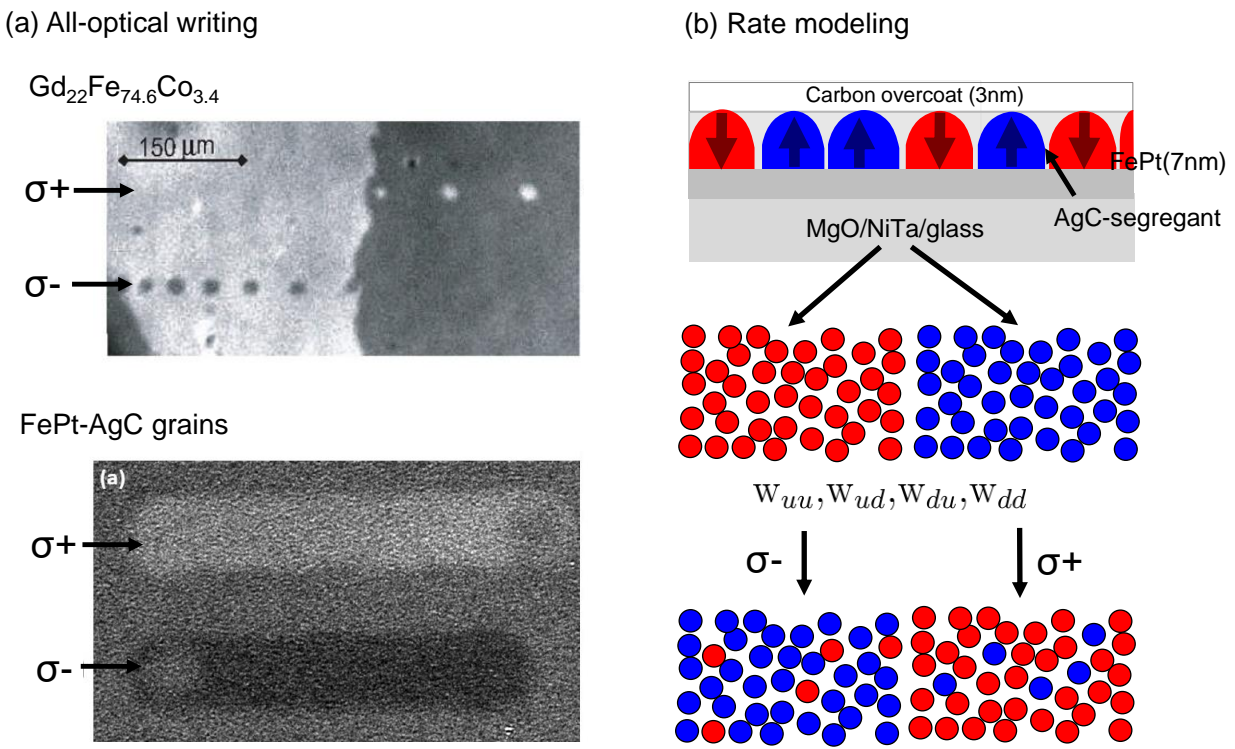

**Fig. 6: All-optical writing.** (a) An asymmetry produces helicity dependence, which can be deterministic, as in rare-earth/transition metals ferrimagnets, or (b) stochastic with probabilities $w_{uu}$, $w_{ud}$, $w_{du}$, $w_{dd}$, for switching in between 'up (u)' and 'down (d)' leading to a final writing rate, here shown as model for nanometer FePt grains in heat assisted recoding media (HAMR). Reprinted from [82,90].

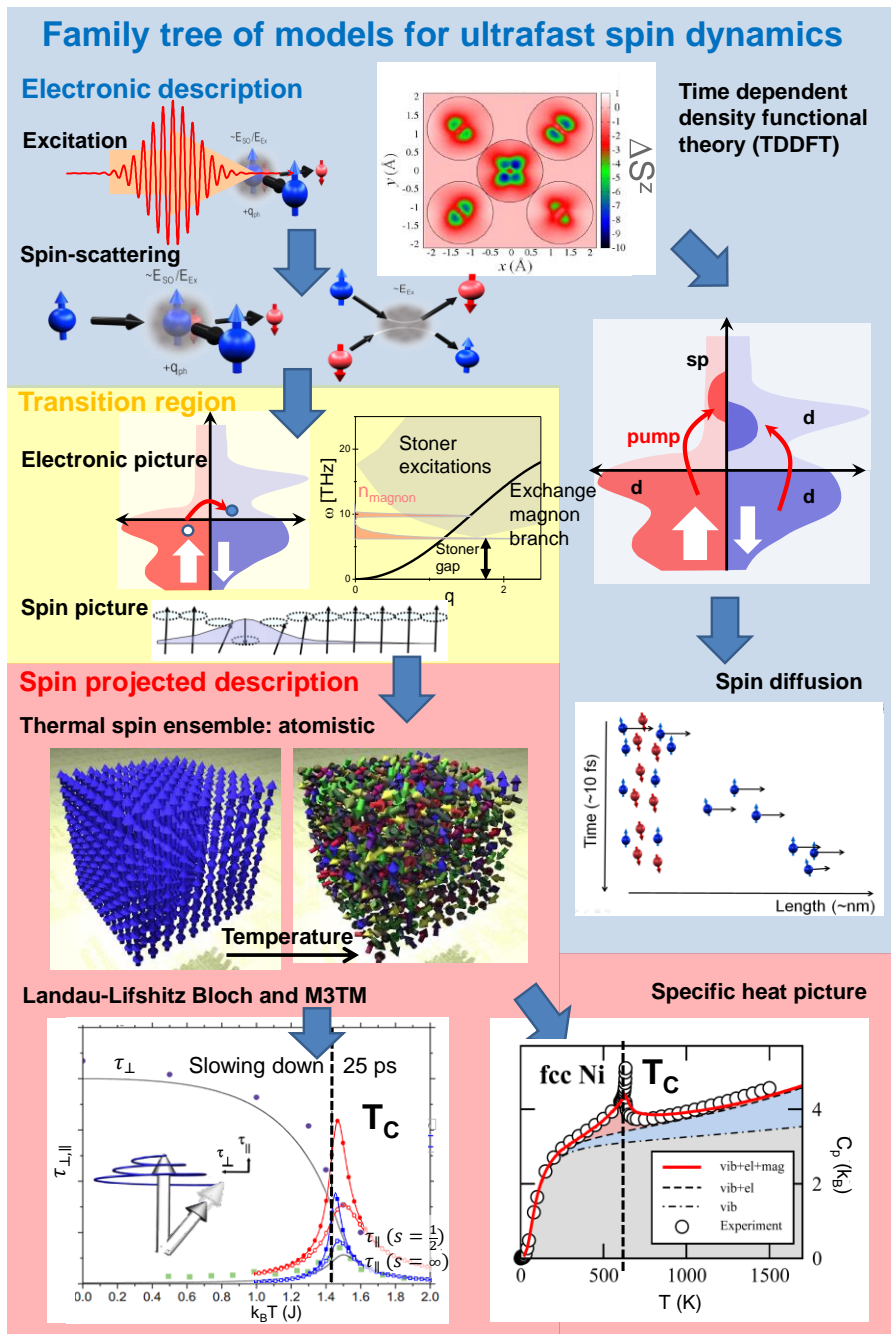

**Fig. 7: Family tree of ultrafast processes and their corresponding description.** The blue area is chosen for predominantly electronic description. The red area for a predominantly spin-based description based on the Heisenberg exchange. The yellow area bridges both regions. Coarse-graining is connected with length and time scales and leads the way to multiscale modeling. Blue area (top): time dependent density functional theory (TDDFT) calculation of the laser driven dynamics for a few atoms in the presence of spin-orbit mixing. Below, microscopic scattering resulting in spin-flip excitations. Yellow area: electronic picture of spin-flip excitations of Stoner type, dispersion and the crossover to the spin picture (high energy exchange spin waves) below. Red area: atomistic modeling of excitation using Langevin fluctuations that can be coarse-grained into the Landau-Lifshitz-Bloch (LLB) description of the thermal response or the microscopic three temperature model (M3TM). Thermal fluctuations, in the same way, are also taken into account in spin specific heat models. Blue area (bottom right): without any scattering or spin-mixing the spin-diffusion models take only into account the no locality of the transport processes. For longer times this has to be extended by including spin-mixing and scattering, as well as other non-local phenomena as spin-momentum transport by spin waves. Parts of this figure are from [69,70,71,72,131,141].

Another source of THz radiation is given by the emission via magnetization dynamics itself. This effect is typically smaller than the electromagnetic field emitted by charge currents because of the pre-factor in the Maxwell equations in front its magnetic terms. The first demonstration of THz emission from magnetization dynamics was reported by magnetization quenching by Bigot et al.[73]. It was recently demonstrated that THz emission can also be used as a probe of the spin-dynamics in antiferromagnets by Kimel et al.[74]. In the first example of THz emitters we used femtosecond lasers to drive the current burst generating the THz emission. Are there also other means to trigger spin-dynamics in the THz range? Spintronic transfer torque devices have been suggested to drive a high frequency spin wave in the THz, schematically shown in Figure 5. To drive this mode efficiently, it must be a coherent mode and have a low damping. An overview over possible material candidates is given in the same figure[75]. It has been demonstrated that ferromagnets exhibiting a large magnetic anisotropy with anisotropy fields around 20 T show coherent precessional dynamics at 0.2-0.6 THz, which opens up the perspective of spin-current driven THz oscillator devices. Using Heusler based materials, these oscillator devices possess a low damping, important for the lifetime and efficiency of the device, as demonstrated by Mizukami et al.[76]. Similar to atomic systems driving coherent light fields, one could imagine THz spin wave lasers driven by this inversion.

**B. All-optical switching**

The milestone that put ultrafast magnetism into application and fueling the whole field, was that of all-optical switching by Rasing, Kimel, Kirilyuk and coworkers[77] discovered in 2007. By the light's polarization, using femtosecond laser pulses of left and right circular polarization, 'up' or 'down' magnetized domain patterns corresponding, to bits of '0' or '1', were written. Today, magnetic memories still store the largest amount of data to date[78]. Analogous to Moore's law, one sees the impressive developments comparing the bit density of 1995 with current memories: there has been a decrease in price by a factor of 100 000. In 1995 the price/GBit in hard disk storage devices reached parity with price of information storage by printing it on paper[79], which set the stage for our information technology and cloud computing today. Strong developments were made in the hard disc industry bringing us TBit/cm memories with bit sizes of ~12×22 nm to a few thousands atoms only over the last years. Large anisotropy materials are needed to work against the thermal fluctuations. This is currently the only way to keep the bit stable for a few years in such small grains of 5-8 nm. Writing fields, which go up to 4 T for current memories, keep the small volume magnetization in plane and the bits '0' and '1' stable at least for five years against Brownian-like thermal fluctuations of the magnetization direction. Schemes to go beyond 10 TBit/in$^2$ are feasible[80]. To overcome large writing fields, heat assisted writing by miniature plasmonic antennas in the read head have been developed to squeeze light into a few nanometer length scales of the actual bits[81]. Employing proper antenna design in future, this procedure should enable the expansion to left- and right circular polarization. Currently, finding a way of deterministic writing using the helicity of light in ferrimagnetic compounds sounds like an elegant way to overcome the writing barrier on the nanoscale.

All-optical writing was demonstrated for ferrimagnetic rare-earth transition metal alloy films GdCoFe or TbCoFe around the compensation temperature of the rare-earth and transition metal moments, Figure 6 (a) shows single shot switching with two helicities[82]. Soon it was found that the situation is quite complex[83],[84]. It became evident that the demagnetization (memory loss of the spin system) together with some symmetry-breaking mechanism determines the physics of the all-optical switching[85]. The magnetization is not reversed by a 180° rotation, as in a standard recording media (transverse relaxation). Due to the heating, the length of the average magnetization is reduced. In the linear switching, a thermal macrospin is represented by the ensemble average moving through zero magnetization (longitudinal relaxation). The decrease of the length of the vector moves the magnetization from +M to –M. In this thermal process, a symmetry-breaking mechanism is needed. In the ferrimagnetic rare-earth transition metal system the two subsystems that are normally antiferromagnetically coupled, form a transient non-equilibrium state with parallel orientation of the moments[86]. This transient state turns the slower rare-earth subsystem with it. The origin of their different electronic and magnetic nature is that the rare-earth systems possess a large magnetic moment and localized 4f states and reacts slowly, versus transition metals with d-states with mixed localized and delocalized character that react quickly. Other symmetry-breaking mechanisms that seed a reversal that are helicity dependent are the inverse Faraday effect[87], inducing a small magnetization of spins and orbital moments in metals[88], and the magneto-optical constants that result in different absorption of left and right helicity of the writing pulse[89], and consequently, a different heating, whose roles in metals is currently being investigated.

As a recent breakthrough, but also leaving many questions and many possibilities, it was shown by Mangin, Fullerton and coworkers[90] that all-optical writing also works for state-of-the-art granular FePt hard disc recording media with high saturation fields of a few Tesla and 5 nm grain size, as shown in Figure 6 (b). In this work, a symmetry-breaking mechanism is present, resulting in a different switching rate for the individual grains that result in switching of the area written by the circularly polarized pump pulse. Switching rates of 30-60% have been reported thus far. Multi-shot experiments suggest that switching rates, $w_{i,j}$, at each laser shot for left and right polarized laser pulses are asymmetric[91], [92]. A certain accumulation is needed to reach the saturation. Two symmetry-breaking effects, the inverse Faraday effect and the different absorption of left and right circularly polarized pulses, can be present. This shows that the observation of optical spin manipulation is very general and that high-density recording media, if parameters reach 100% switching rates, could be written all-optically in the far future.

**III. Theoretical perspectives**

Optimization of ultrafast magnetic devices is only possible by theoretical understanding. From the outset of this section, it has to be clarified: there is no unique model of ultrafast magnetization dynamics. The family of models on different timescales is depicted in Figure 7, showing the complexity of the processes involved on different length and time scales. One can roughly distinguish ultrafast electronic processes driven by the light field and non-equilibrium electron distributions that act on the spin state, shown as the blue area in Figure 7. We switch from an electronic description to a stochastic spin-ensemble, connected by the yellow area in Figure 7. Internal local-field fluctuations connect these initial electronic dynamics driven by the laser field to a spin-model and finally to stochastic spin fluctuations arising from spin scattering processes. Further coarse-graining permits the step from individual spins in an atomistic description to a macroscopic description of ultrafast phase transitions and macroscopic magnetization dynamics, depicted as the red area in Figure 7. Non-locality of the excitation also triggers ultrafast transport of spin currents and THz spin waves, shown by the blue area. Finally, local and non-local processes have to equilibrate in the long term and have to be connected. In a full description they cannot be independent of each other. One of the main players is the spin-orbit interaction, resulting in spin-mixing spin-orbit fields, and transport effects on picometer (intra-atomic), to nanometer (interfaces) and micron (gradients) length scales. Without spin-orbit interaction, no demagnetization at all would be observed. Processes can be classified in the family tree of theoretical descriptions of ultrafast magnetization dynamics, divided into three areas:

  *(i) Electronic description*: electronic processes, light driving the atomic electrons (coherent), dynamic density functional theory, dynamic mean field theory (DMFT). Delocalized spin models: spin currents though interfaces, ballistic to diffusive.

  *(ii) Transition region:* spin-dynamics interfacing (i) and (iii). Elliott-Yafet scattering, electron magnon interactions and Stoner excitations, bridging the electronic processes to the spin-separated description and atomistic spin models.

  *(iii) Spin projected description:* local spin models: atomistic modeling with thermal fluctuations, stochastic macroscopic ensemble (Landau-Lifshitz-Bloch (LLB)), microscopic three temperature model (M3TM), spin specific heat.

Rather than to give a complete survey on the theoretical methods, the aim of this part of the perspective is to describe from an experimentalist's point of view how different approaches interact on a multiscale level. To start, the smallest length scale and fastest time scale has to start with light driven currents within the atom. The current development of time dependent density functional theory (TDDFT) is now able to resolve the laser driven spin-dynamics in an ensemble of a few ferromagnetic atoms[93,94], or even meanwhile in Heusler compounds[95]. The method is able to map what happens in the laser field driven system. It mirrors the role of the spin-orbit coupling and can depict where around an atom the reduction of magnetization is the strongest, e.g. close to the center (Figure 7, top). These atomic currents excited coherently with the laser field are still maintained shortly after the pulse and result in an evolution of spin and momentum, decreasing the magnetization.

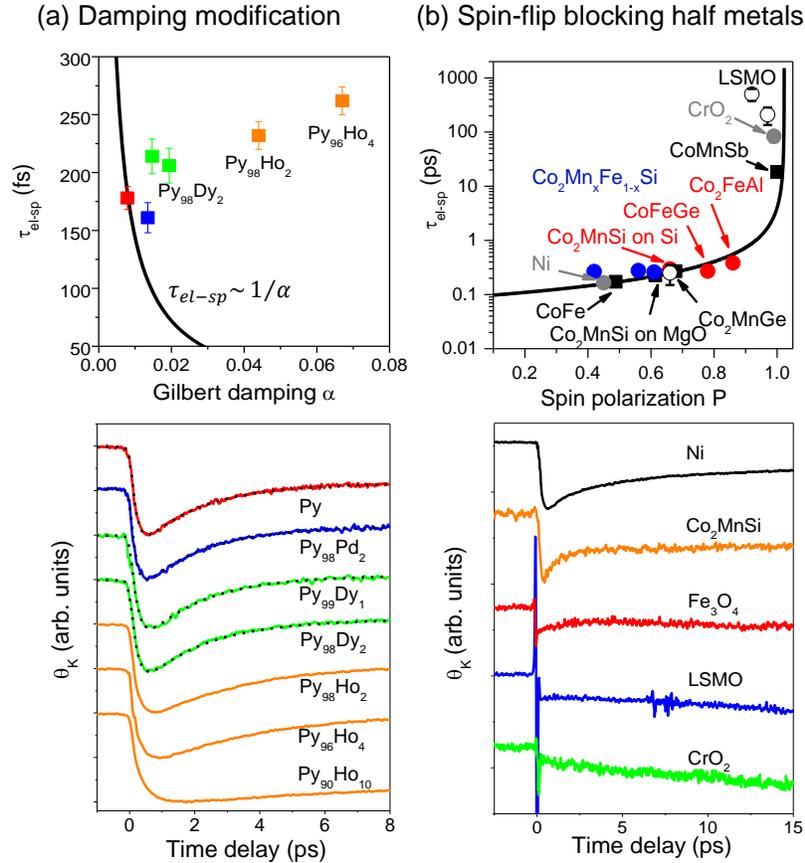

**Fig. 8: Experimental tuning of the spin scattering channels**. (a) Relation of Gilbert damping $\alpha$ and ultrafast demagnetization for rare-earth doping, data taken from Walowski et al.[106]. While a large Gilbert damping should be associated with a fast demagnetization $\tau_{el-sp} \sim 1/\alpha$, rare-earth metals behave opposite because of the slow relaxation impurity model (localized 4f states). (b) Decreasing the spin scattering channels for large (La$_{0.67}$Sr$_{0.33}$MnO$_3$ (LSMO), CrO$_2$) and small (Heusler compounds) half metallic gaps resulting is a blocking of the spin scattering channels. The solid line is given by $\tau_{el-sp} \sim \tau_{el,0}/c^2(1-P)$, where P is the spin polarization at the Fermi level and c the spin-mixing parameter. Data taken from Steil et al., Müller et al. and Mann et al. [108,109,110,111,112].

On the next level, relativistic band structure calculations for a full crystal can give an insight in the change of the spin and orbit quantum number as the electron propagates. The spin-mixing can be calculated, which gives the probability of a spin-flip, as the electron propagates in the bands and the energy and momentum are changed, especially in case of a scattering event. However, this is not the only source of spin-flips. One way to connect the electron excitation with the spin system has been described by Elliott[96,97,98]. While the spin-flip arising specifically from electron-phonon scattering was discussed in the spin relaxation paramagnetic metals by Yafet and called Elliott-Yafet scattering later[99], in the original work by Elliott the spin-mixing is described much more general, yielding some confusion in the community. Conduction electron spin resonance (CESR) is very important to understand the spin relaxation of the electrons in metals. This allowed to study the temperature dependent spin relaxation of conduction electrons via the line width and gave the first insights into the time scales of spin relaxation[100,101,102]. It motivated the theoretical work. Later the Kamberskyy model[103] was developed for energy relaxation of the ferromagnetic resonance (FMR). Near the Fermi level, the spin's precession in the FMR experiment has some effect on the electronic structure and spin scattering events occur, leading to dissipation. This breathing and wobbling of the Fermi surface, corresponds to intraband and interband transitions, in the tail of the Fermi distribution (~25 meV energy window at room temperature). This leads to magnetic damping, called Gilbert damping $\alpha$. Although this is related to the spin-flips at ultrafast time scales, for an excited electron system it is clear that the situation is more complex and more bands will be involved. The main difficulty here is to describe the different scattering events in a realistic model and the implementation of the phonons involved[98]. Simplifications can be made to separate localized bands and delocalized bands, possessing different spin scattering rates and spin polarization[104], as it was also proposed as a spin transport model by Stearns[105]. Again, this shows how the spin-dynamics picture is connected to questions of spin transport in general.

This connection of energy dissipation and microscopic spin scattering has been tested in different model systems experimentally. Are changes of the electronic structure at the Fermi level resulting in a larger Gilbert damping $\alpha$ connected to a change in the energy dissipation on the ultrafast time scale, e.g. faster demagnetization because of a stronger coupling? Experiments by Walowski et al.[106], similar results have been found by and Radu et al.[107], are summarized in Figure 8 (a): rare-earth doping of a ferromagnet is known to increase the magnetic damping. However, surprisingly on ultrafast time scales, the demagnetization does not get faster, it gets even slower. One reason is that for rare-earth materials the damping comes from the energy transfer at the 4f levels. These are not located at the Fermi level and the mechanism of energy transfer is slow. Another test case for the connection of microscopic and macroscopic spin scattering are half metals. The spin-flip scattering can be turned off by the half metallic character of the material, e.g. Figure 3 (b), resulting in a decoupling. This has been tested in a series of half metals, oxides and Heusler alloys with high spin-polarization at around the Fermi level, a summary of Müller et al.,[108] Steil et al.[109,110,111] and Mann et al.[112] is given in Figure 8 (b). While it seems that there is the expected correspondence of demagnetization becoming slower at high spin polarizations that would be consistent with a spin scattering scenario, especially for the small band gap Heusler alloys with specific positions of the Fermi level close to a band edge, these effects can also be largely suppressed due to electrons (or holes) excited to higher energies above the gap. Future understanding of these processes can only be generated from such experiments controlling the materials properties. In our view, much more information is needed from experiment for different materials to trace origin of this connection.

These spin-flips are connected via Stoner excitations[113,114] and decay into spin waves heating the magnetic system, driving it on ultrashort time scales to temperatures close to the Curie temperature, $T_C$, or even above (Figure 7, middle). In driving a ferromagnet though the phase transition, spin-fluctuations are become so large that the averaged magnetization breaks down even in presence of the exchange interaction. The exchange interaction is unchanged and vanishes at temperatures when the thermal energy is equal to $kT \sim J_{ex}$, which is much larger than $T_C$. The presence of fluctuations is a typical signature of a phase transition, where diverging time and length scales result in a slowing down of the response of the system. It results in a strong increase of the spin specific heat at around the phase transition. One has to keep in mind that in the high-density limit of spin waves (region $T_C > T > T_C/2$) the spin system is not described by weakly-interacting spin waves, rather by spin-fluctuations of strongly coupled spin waves with higher order interaction. This allows the creation of new metastable phases in magnetic materials by ultrafast quenching of this excited state as complex vortex networks or magnetic skyrmion bubbles, and allows one to study the physics of phase transition phenomena from different perspectives, while these particle-like solitons are 'born'.

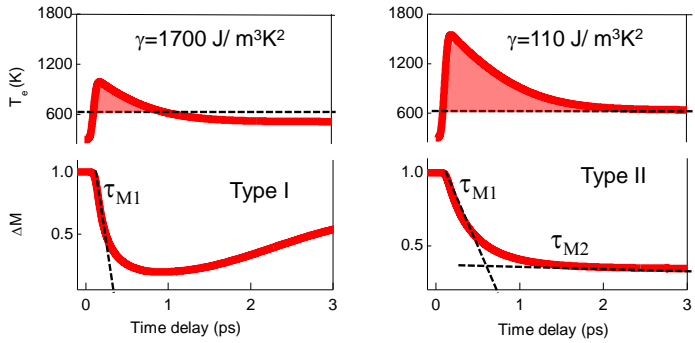
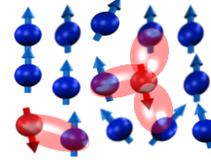
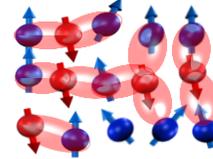
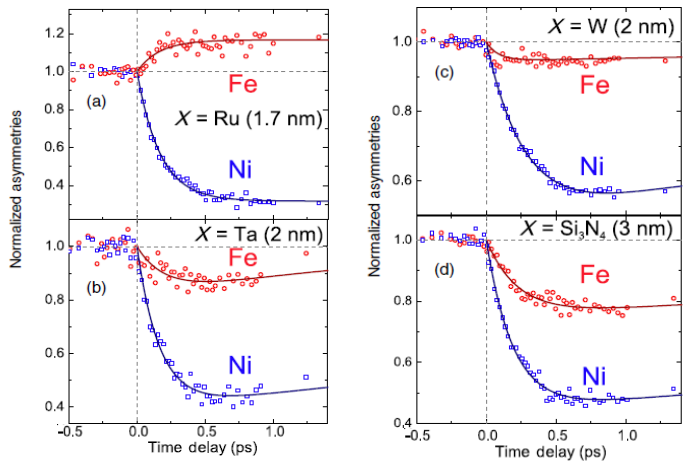
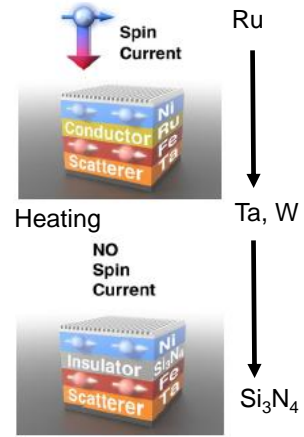

**Fig 9: Traces of critical phenomena and spin transport in ultrafast demagnetization experiments**. (a) Comparison of two scenarios for Fe and FePt (large and small Sommerfeld constant $\gamma$), showing critical behavior (slowing down) and how it is related to the electron temperature (described with the LLB equation). (b) Control of aspects of spin-diffusion by the choice of the separation layer from transparent, to absorbing or insulating. For the case of Ru spacer layer the spin current effects dominate the ultrafast dynamics. For the case of a spin scatterer or insulator only heating is observed. Reprinted from [115,116].

While the spin-flip mechanisms connect the electronic description with the spin-projected description, more in detail discussed the following section, the transport character can be described in the electronic picture taking specific velocities and relaxation rates for the electrons in their spin separated bands. Lateral distances in the nanometer range, connecting ferromagnets with a non-metallic interface, is leading to a lateral redistribution of magnetization as described by Battiato et al.[34]. Typically majority spins are more mobile, depicted in Figure 7 on the right. As a consequence, it leads to an ultrafast demagnetization of the films. However, it can also lead to an increase of the magnetization if the spin-currents are injected into a second ferromagnetic layer with the same orientation of the majority spin, see Figure 9 (b). The direct detection of these currents was described in II A "THz spintronics" more in detail in this context. Two further models that connect to the transport properties, a shift of the chemical potential and relaxation in between the different bands (s-d relaxation), have been proposed by Rethfeld[117] and Manchon[118].

## A. Coarse-grained thermal model: A perspective for predictive power

Similar to micromagnetic models gaining high predictive power and becoming an indispensable tool for nanomagnetism[119] in the last decade, thermal models, meanwhile, reach predictive power for ultrafast experiments[120]. For example, it will be possible to optimize writing asymmetries in all-optical writing experiments to reduce fluence thresholds in the future. The role of spin wave or spin-cluster fluctuations in the characteristic response of the ferromagnet has been suggested very early in 2007 in parallel by different groups by atomistic[121], thermal macrospin[122,123], or simple micromagnetic spin-fluctuation[124] models. On one hand, the time scales in the demagnetization experiments were found to be fluence dependent when reaching strong demagnetization beyond 50% of the saturation magnetization. Therefore, there was a need to include spin-fluctuations. On the other hand, when simulating an excited spin system using thermal noise mimicking the fluctuations, the importance of high energy THz spin waves for the remagnetization processes had been observed.

The different thermal coarse-grained models in this section are often quoted as phenomenological, however, they are derived from very basic foundations and have, if used in a suited way, predictive power. They can, in principle, all be derived *ab initio*. We will compare the different coarse-grained models in the following. All thermal models, coupling an electron temperature to a spin system via an electron-spin relaxation rate, are very similar because they have the same physical ingredients. The most fundamental thermal spin model is the direct simulation atomistic spins[125,126], however, this model is limited to small spin ensembles of $10^6$ spins. Parameters of the exchange and magnetic moments can be taken from *ab initio* calculations or adapted to correctly describe the equilibrium behavior of the magnetization which can easily be experimentally verified and gives a test of the model. The electron-spin relaxation is the connecting factor for the temperature in the Langevin dynamics and can be taken from experiments. It is determined by the measurement of the magnetic Gilbert damping $\alpha$. This factor, in the macroscopic ensemble becomes temperature dependent and a new relaxation time for restoring the magnetization arises that becomes important when approaching $T_C$. Very critical for all the models is the input of the temperature of the electron system that is coupled by the Langevin dynamics. It is also possible to take the experimental temperature and extract the corresponding parameters directly from the reflectivity dynamics and get closest to the experimental temperature values. Historically, the first thermal model developed was the thermal macrospin model from Garanin[127], initially thought to describe temperature effect on spin systems. It is derived from an atomistic form of the Landau-Lifshitz-Gilbert equation of a spin ensemble in the presence of a thermal Langevin field[128]. Garanin showed [129,130] that the equations for this ensemble can be transformed to an equation of motion for a single macrospin where, in order to account for the thermal dynamics, a second Bloch-like relaxation term is found. Again, we get two relaxations: the transverse relaxation and the longitudinal Bloch-like relaxation, therefore called Landau-Lifshitz-Bloch (LLB) model[131,132,133]. It was shown that the atomistic and the thermal macrospin model give the same results to describe the thermal equilibrium magnetization and damping[134]. Important are the magnetic susceptibilities that describe the response times to heat stimulus, which includes the slowing down near the phase transition, seen in Figure 9 (a) for a FePt film. The slowing down phenomena is typical for the phase transition.

In this thermal macrospin model, one uses a thermal noise term for a spin system with many spins[135]. By averaging this thermal ensemble, the total length of a thermal spin will decrease. The exchange interaction is always present locally. However thermal fluctuations drive the system macroscopically paramagnetic at the phase transition in the mean field by disorder. With the Landau-Lifshitz-Bloch model, complex magnetization dynamics can be described, due to the inclusion of Landau-Lifshitz-Gilbert type dynamics, and can be numerically implemented into standard micromagnetic models. This description is very versatile in describing all-thermal effects in micromagnetism, and goes beyond the computational and numerical complexity of an atomistic calculation[136]. A relative to the LLB model is the microscopic three temperature model (M3TM)[137]. It was developed specifically for ultrafast demagnetization, its equations are therefore comparably simple and it is often implemented. M3TM considers a longitudinal relaxation only, calculated from the slope of the M(T) curve, and the phonon system is treated within this model in the Einstein model. It was shown that the M3TM and LLB model are equal in the description of the heat-triggered, longitudinal spin-dynamics[138]. In certain limits both can be solved analytically, and novel scenarios can be developed within these frameworks.

A third suggestion had been made by Kimling et al.[139], using the spin specific heat as it diverges at around $T_C$ itself in the modeling. While, in principle, spin-fluctuations are captured in the LLB and M3TM models via the slope of M(T) that enters as a slowing down of the spin system's response via the dynamics susceptibility, there is no feedback loop for the spin system to the electrons or phonons. Using the spin specific heat that is connected to the magnetization M(T) via $c_{sp} \sim \partial M^2/\partial T$ (in mean field approximation)[140] that means: if the M(T) curve becomes steep, the spin specific heat increases. This directly compares with the increase in the dynamic susceptibility and the longitudinal relaxation time, shown at the bottom of Figure 7, side by side, and is responsible the for the slowing down for strong demagnetization reaching T ~ $T_C$. In the LLB equations this feedback on the total specific heat is not captured and has to be specifically incorporated. It can become relevant if the dynamics of the temperature of the electronic system is not dominated by the electron-phonon coupling, e.g. for systems where $G_{el-ph}$ is small. Two examples of experimental data and the theoretical description using the M3TM or the spin heat capacity are shown in Figure 10 for a series of pump-probe experiments heating the base temperature of the sample to $T_C$. The ultrafast demagnetization goes from a peak-like structure, from a fast response of the spin system, to a slow step-like decrease of the magnetization (increase of the spin-temperature). When the fluctuations become dominating (spin specific heat increases), the spin system cannot follow the electron temperature anymore. This would be normally the case when the spin specific heat is negligible. It can amount to one third of the total heat capacity near $T_C$; calculations taken from a detailed topical review by Hickel et al. [141] are given at the bottom of Figure 7 on the right. Other approaches in the family of thermal models are the Baryakhtar equation[142] especially suited for multisublattice systems, and the self-consistent Bloch (SCB) equation[143].

Different approaches investigate to couple thermal model to models of spin-diffusion, e.g. the electronic and the spin-projected description[144]. Experiments shown in Figure 9 (b) by Turgut et al.[116] demonstrate that depending on the interlayer, spin current and thermal effects can be turned on and off. In addition for rare earth materials dynamic band structure effects may have to be included in future models[145].

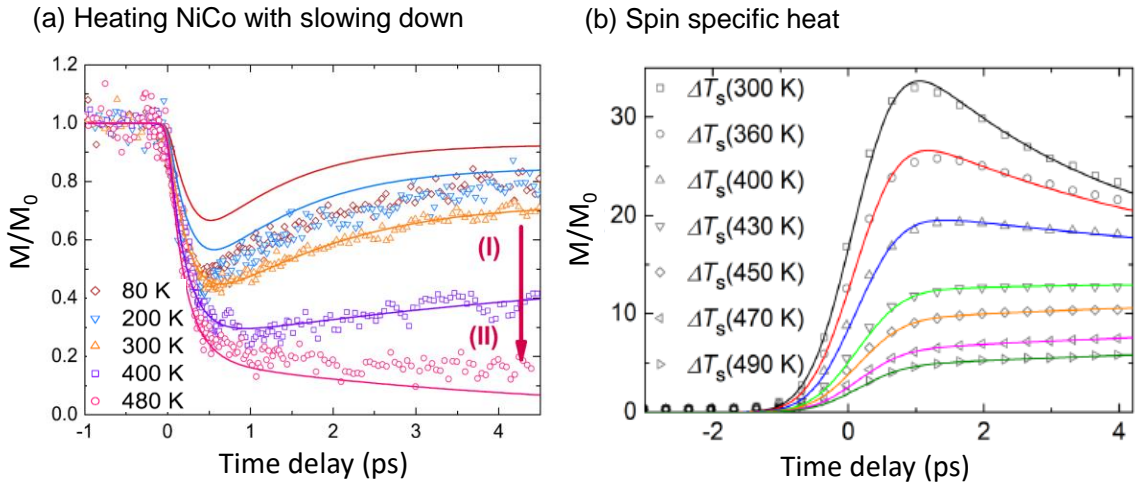

**Fig. 10: Manifestation of spin-fluctuations in the experiments and approaches to analyze the dynamics based on thermal models**. (a) Response changing at around $T_C$ modeled in the M3TM or in (b) using the spin specific heat itself for a FePt-Cu alloy, reprinted from [[146],[139]].

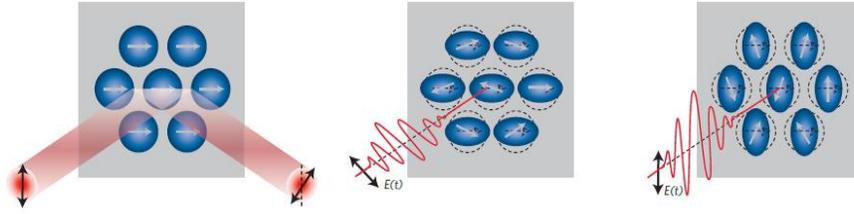

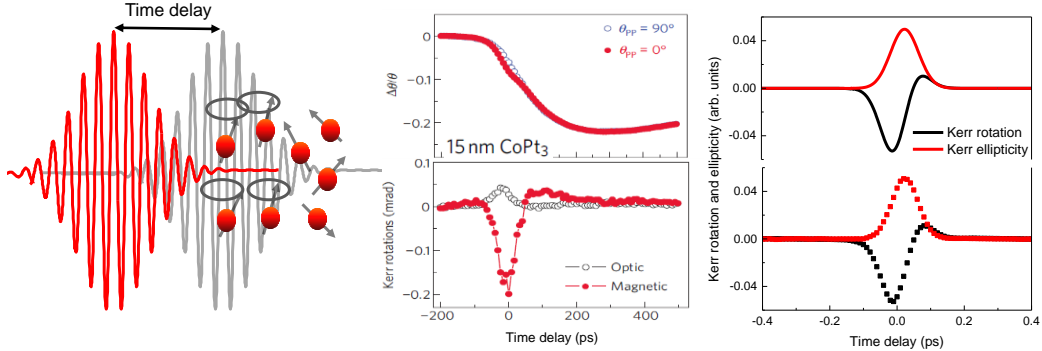

**Fig. 11: Coherent control.** In ferromagnets and topological insulators, from [150,154,147].

## IV. Ultimate timescale: The Future of coherent control

The ultimate way to gain control over magnetism is through coherent excitation with a light field[148]. This implies an interaction of the laser field directly with the spin system. While coherent control seems feasible with ultrastrong THz field pulses, where the B-field amplitude reaches the Tesla range, there are reports that too much heat is deposited and the coherence is disturbed[149]. For light in the visible region, coherent excitation of ferromagnetism and a corresponding model has been proposed by Bigot et al.[150]. In this detailed experiment they extracted coherent signals that are only present as the laser pulse interacts with the sample, presented in Figure 11, for a $CoPt_3$ film. One can picture a polarization that is driven by the light in a transient state. Those ultrashort polarization effects are also known from other material systems such as MnGaAs[151] and manganites[152]. They leave a typical fingerprint in the complex Kerr rotation that can be described in a Raman-type model. Other approaches have been developed for metals[153]. An interesting pathway is to use this coherent polarization to trigger interactions with another part of the magnetic subsystem as for example the spin-polarized surface states in topological insulators, as seen in the different response for the components of the complex Kerr rotation from the $Bi_2Se_3$ family, $(Bi_{0.57}Sb_{0.43})_2Te_3$ presented in Figure 11 (b)[154]. The hope is that these processes are faster than the thermal demagnetization effect. Their investigation will shed light on the inverse Faraday effects and further ultrafast processes that happen faster than the scattering time of the electrons in a coherent state, ultimately leading to attosecond control[155] of magnetization.

## V. Conclusion

In this perspective, we have shown that ultrafast magnetism has arrived at the stage of quantitative prediction and understanding. Modeling will become an important aspect for predictions: the understanding how much power can be saved for all-optical writing to make it efficient within multiscale approaches, leads to new ultrafast all-optical nanomemories addressing nanometer FePt grains. On all timescales, the spin-orbit interaction is one of the main players acting in two ways: resulting in switching asymmetries via magnetic-optics and the control of spin-flips. On the other hand, spin-orbit effects and spin-dependent transport can be controlled on THz time scales for applications. Ultrafast laser pulse based trigger and control of the spin currents and ultrafast spin waves set the stage for THz spintronics. We believe that the combination of ultrafast magnetism and spintronics will have more interesting discoveries in fundamental physics and applications to come.

**Acknowledgements**

M.M. thanks the German research foundation (DFG) for support within the projects MU 1780/ 6-1 Photo-Magnonics: Materials and Devices, MU 1780/8-1,2 in the priority program "Spincaloritronic Transport" (SPP 1538) and MU 1780/10-1,2 in the priority program "Topological Insulators: Materials - Fundamental Properties – Devices" (SPP 1666).

**References**

[1] E. Beaurepaire, J.-C. Merle, A. Daunois, and J.-Y. Bigot, Phys. Rev. Lett. **76**, 4250 (1996).
[2] A. Kirilyuk, A. V. Kimel, and T. Rasing, Rev. Mod. Phys. **82**, 2731 (2010).
[3] J.-Y. Bigot and M. Vomir, Annalen der Physik **525**, 2 (2013).
[4] C. Stamm, et al., Nature Mater. **6**, 740 (2007).
[5] C. Boeglin, et al., Nature **465**, 458 (2010).
[6] B. Vodungbo, et al., Nature Comm. **3**, 999 (2012).
[7] C. La-O-Vorakiat, et al., Phys. Rev. Lett. **103**, 257402 (2009).
[8] C. E. Graves, et al. Nature **12**, 3597 (2013).
[9] A. Eschenlohr, M. Battiato, P. Maldonado, N. Pontius, T. Kachel, K. Holldack, R. Mitzner, A. Fröhlisch, P. M. Oppeneer, and C. Stamm, Nature **12**, 3546 (2013).
[10] B. Frietsch, J. Bowlan, R. Carley, M. Teichmann, S. Wienholdt, D. Hinzke, and U. Nowak, K. Carva, P. M. Oppeneer, and M. Weinelt. Nature Comm. **6**, 8262 (2015).
[11] H.-S. Rhie, H.A. Dürr, W. Eberhardt, Phys. Rev. Lett. **90**, 247201 (2003).
[12] E. Carpene, E. Mancini, C. Dallera, M. Brenna, E. Puppin, and S. De Silvestri, Phys. Rev. B **78**, 174422 (2008).
[13] E. Carpene, F. Boschini, H. Hedayat, C. Piovera, and C. Dallera, E. Puppin, M. Mansurova, M. Münzenberg, X. Zhang and A. Gupta, Phys. Rev. B 87 (2013).
[14] B. Koopmans, J. J. M. Ruigrok, F. Dalla Longa, and W. J. M. de Jonge, Phys. Rev. Lett. **95**, 267207 (2005).
[15] W. Heisenberg, Zeitschrift für Physik **38**, 411 (1926).
[16] P. A. M. Dirac, Proc. R. Soc. A **112**, 661 (1926).
[17] E. Stoner, Proc. R. Soc. A **154**, 565 (1936); O. Gunnarson, J. Phys. F: Metal. Phys. **6** (1976).
[18] R. F. L. Evans, W. J. Fan, P. Chureemart, T. A. Ostler, M. O. A. Ellis, R. W. Chantrell, J. Phys.: Condens. Matter **26**, 103202 (2014).
[19] T. L. Gilbert, IEEE Trans. Magn. **40**, 3443 (2004); M. Fähnle et al., Phys. Rev. B **73**, 172408 (2006).
[20] M. I. D'yakonov and V. I. Perel, Phys. Lett. **35**, 459 (1971); J. E. Hirsch, Phys. Rev. Lett. **83**, 1834 (1999); S. Zhang, Phys. Rev. Lett. **85**, 393 (2000).
[21] M. Gradhand, M. Czerner, D. V. Fedorov, P. Zahn, B. Yu. Yavorsky, L. Szunyogh, and I. Mertig, Phys. Rev. B **80**, 224413 (2009).
[22] The degree of mixture is given by the exchange interaction separating the spin-split bands over its counterpart, the spin-orbit interaction, mixing them and are different at each k-point even in the same band.

[23] N. H. Long, P. Mavropoulos, S. Heers, B. Zimmermann, Y. Mokrousov, and S. Blügel, Phys. Rev. B **88**, 144408 (2013).
[24] D. Steiauf and M. Fähnle, Phys. Rev. B **79**, 140401(R) (2009); C. Illg, M. Haag, and M. Fähnle, Phys. Rev. B **88**, 214404 (2013).
[25] B. Andres, M. Christ, C. Gahl, J. Kirschner, M. Wietstruk, and M. Weinelt, Phys. Rev. Lett. **115**, 207404 (2015).
[26] M. Born, J. R. Oppenheimer, Annalen der Physik **389**, 457 (1927)
[27] F. Dalla Longa, J. T. Kohlhepp, W. J. M. de Jonge, and B. Koopmans, Phys. Rev. B **75**, 224431 (2007).
[28] T. F. Nova, A. Cartella, A. Cantaluppi, M. Foerst, D. Bossini, R. V. Mikhaylovskiy, A. V. Kimel, R. Merlin, A. Cavalleri, arXiv:1512.06351; T. Kampfrath et al. unpublished.
[29] D. Steiauf, C. Illg and M. Fähnle, J. Phys.: Conf. Ser. **200**, 042024 (2010).
[30] N. F. Mott, Proc. R. Soc. Lond. A **153**, 699717 (1936).
[31] T. Valet, and A. Fert, Phys. Rev. B **48**, 7099 (1993).
[32] M. N. Baibich, J. M. Broto, A. Fert, F. Nguyen Van Dau, F. Petroff, P. Etienne, G. Creuzet, A. Friederich and J. Chazelas, Phys. Rev. Lett. **61**, 2472 (1988).
[33] J. Barnaś, A. Fuss, R. E. Camley, P. Grünberg, and W. Zinn, Phys. Rev. B **42**, 8110 (1990)
[34] M. Battiato, K. Carva, and P. M. Oppeneer, Phys. Rev. Lett. **105**, 027203 (2010)
[35] A. Melnikov, I. Razdolski, T.O. Wehling, E.Th. Papaioannou, V. Roddatis, P. Fumagalli, O. Aktsipetrov, A.I. Lichtenstein, U. Bovensiepen, Phys. Rev. Lett. **107**, 076601 (2011).
[36] G. E. W. Bauer, E. Saitoh, B. J. van Wees, Nature Mater. **11**, 391 (2012).
[37] M. Battiato, K. Carva, and P. M. Oppeneer, Phys. Rev. B **86**, 024404 (2012).
[38] C. Gyung-Min, B.-C. Min, K.-J. Lee, and D.G. Cahill, Nature Comm. **5**, 4334 (2013).
[39] S. Kaltenborn, H. C. Schneider in Ultrafast Magnetism I, Volume 159, Springer Proceedings in Physics, 169-171(2014).
[40] R. Metzler and J. Klafter, Phys. Rep. **339**, 1 (2000).
[41] D. Brockmann und T. Geisel, Phys. Rev. Lett. **90**, 170601 (2003).
[42] A. Slachter, F. L. Bakker, J. P. Adam and B. J. van Wees, Nature Phys. **6**, 879 (2010).
[43] K. Uchida, H. Adachi, T. Ota, H. Nakayama, S. Maekawa and E. Saitoh, Appl. Phys. Lett. **97**, 172505 (2010).
[44] G. Tveten, A. Brataas, and Y. Tserkovnyak, Phys. Rev. B **92**, 180412(R) (2015).
[45] A. Hoffmann and S. D. Bader, Phys. Rev. Appl. **4**, 047001 (2015).
[46] Editorial, Nature Nanotech. **10**, 185 (2015).
[47] https://www.everspin.com/
[48] H.-S. Philip Wong and S. Salahuddin, Nature Nanotech. **10**, 191 (2015).
[49] B. Lenk, H. Ulrichs, F. Garbs, M. Münzenberg, Phys. Rep. **507**, 107 (2011).
[50] International Technology Roadmap for Semiconductors (ITRS), http://public.itrs.net/
[51] T. Kampfrath, K. Tanaka and K. A. Nelson, Nature Photon. **7**, 680 (2013); Editorial, Nature Photon. **7**, 665 (2013).
[52] G.-X. Miao, M. Münzenberg and J. S. Moodera, Rep. Prog. Phys. **74**, 036501 (2011).
[53] Z. Jin, A. Tkach, F. Casper, V. Spetter, H. Grimm, A. Thomas, T. Kampfrath, M. Bonn, M. Kläui and D. Turchinovich, Nature Phys. **11**, 761 (2015).
[54] A. J. Schellekens, K. C. Kuiper, R. R. J. C. de Wit, and B. Koopmans, Nature Comm. **5**, 4333 (2014).
[55] M. Savoini, C. Piovera, C. Rinaldi, E. Albisetti, D. Petti, A. R. Khorsand, L. Duò, C. Dallera, M. Cantoni, R. Bertacco, M. Finazzi, E. Carpene, A. V. Kimel, A. Kirilyuk, and Th. Rasing, Phys. Rev. B **89**, 140402(R) (2014).
[56] J. C. Leutenantsmeyer, M. Walter, V. Zbarsky, M. Münzenberg, R. Gareev, K. Rott, A. Thomas, G. Reiss, P. Peretzki, H. Schuhmann, M. Seibt, M. Czerner, C. Heiliger, SPIN **3**, 1350002 (2013).
[57] M. Walter, J. Walowski, V. Zbarsky, M. Münzenberg, M. Schäfers, D. Ebke, G. Reiss, A. Thomas, P. Peretzki, M. Seibt, J. S. Moodera, M. Czerner, M. Bachmann, C. Heiliger, Nature Mater. **10**, 742 (2011).

58 A. Pushp, T. Phung, C. Rettner, B. P. Hughes, S.-H. Yang, and S. S. P. Parkin, PNAS **112**, 6585 (2015).
59 W. He, T. Zhu, X.-Q. Zhang, H.-T. Yang and Z.-H. Cheng, Sci. Rep. **3**, 2883 (2013).
60 A. Manchon and S. Zhang, Phys. Rev. B **78**, 212405 (2008).
61 L. Liu, O. J. Lee, T. J. Gudmundsen, D. C. Ralph, and R. A. Buhrman, Science **336**, 555 (2012).
62 I. M. Miron, G. Gaudin, S. Auffret, B. Rodmacq, A. Schuhl, S. Pizzini, J. Vogel, and P. Gambardella, et al., Nature Mater. **9**, 230 (2010).
63 T. Seifert, S. Jaiswal, U. Martens, J. Hannegan, L. Braun, P. Maldonado, F. Freimuth, A. Kronenberg, J. Henrizi, I. Radu, E. Beaurepaire, Y. Mokrousov, P. M. Oppeneer, M. Jourdan, G. Jakob, D. Turchinovich, L. M. Hayden, M. Wolf, M. Münzenberg, M. Kläui, T. Kampfrath, Nature Photon., advance online publication, doi:10.1038/nphoton.2016.91.
64 T. Kampfrath, M. Battiato, P. Maldonado, G. Eilers, J. Nötzold, S. Mährlein, V. Zbarsky, F. Freimuth, Y. Mokrousov, S. Blügel, M. Wolf, I. Radu, P. M. Oppeneer. and M. Münzenberg, Nature Nanotech. **8**, 256 (2013)
65 M. Tonouchi, Nature Photon. **1**, 97 (2007); H. Siegel, IEEE MTT **50**, 910 (2002).
66 Y. Kawano, and K. Ishibashi, Nature Photon. **2**, 618 (2008).
67 A. J. Huber, F. Keilmann, J. Wittborn, J. Aizpurua, and R. Hillenbrand, Nano Lett. **8**, 3766 (2008).
68 T. J. Huisman, R. V. Mikhaylovskiy, J. D. Costa, F. Freimuth, E. Paz, J. Ventura, P. P. Freitas, S. Blügel, Y. Mokrousov, Th. Rasing and A. V. Kimel, Nature Nanotech. **11**, 455–458 (2016).
69 J. Simoni, M. Stamenova, S. Sanvito, arXiv:1604.06262.
70 Atomistic calculation and image by Ulrich Nowak University of Konstanz.
71 N. Kazantseva. et al. Phys. Rev. B **77**, 184428 (2008).
72 O. Chubykalo-Fesenko, U. Nowak, R. W. Chantrell, and D. Garanin, Phys. Rev. B **74**, 094436 (2006).
73 E. Beaurepaire et al., Appl. Phys. Lett. **84**, 3465 (2004).
74 V. Mikhaylovskiy, E. Hendry, A. Secchi, J. H. Mentink, M. Eckstein, A. Wu, R. V. Pisarev, V. V. Kruglyak, M. I. Katsnelson, T. Rasing and A. V. Kimel, Nature Comm. **6**, 8190 (2015).
75 S. Mizukami, A. Sakuma, A. Sugihara, K.Z. Suzuki, R. Ranjbar, Scripta Materiaria **118**, 70 (2016).
76 S Mizukami, A Sugihara, S Iihama, Y Sasaki, KZ Suzuki, T Miyazaki, Appl. Phys. Lett. **108**, 012404 (2016).
77 C. D. Stanciu, F. Hansteen, A. V. Kimel, A. Kirilyuk, A. Tsukamoto, A. Itoh, and Th. Rasing, Phys. Rev. Lett. 99, 047601 (2007).
78 Ultra-High-Density Magnetic Recording: Storage Materials and Media Designs, editors G. Varvaro, F. Casoli, Pan Stanford (2016).
79 Materials Research Society Bulletin 31, 5 May 2006 (2006).
80 C. Vogler, C. Abert, F. Bruckner, and D. Suess, Appl. Phys. Lett. **108**, 102406 (2016).
81 B. C. Stipe, et al., Nature Photon. **4**, 484 (2010).
82 C. D. Stanciu, F. Hansteen, A.V. Kimel, A. Kirilyuk, A. Tsukamoto, A. Itoh, and Th. Rasing. Phys. Rev. Lett. **99**, 047601, (2007).
83 B. Hebler, A. Hassdenteufel, P. Reinhardt, H. Karl and M. Albrecht, Frontiers in Materials 3, 8 (2016).
84 S. Mangin, M. Gottwald, C-H. Lambert, D. Steil, V. Uhlír, L. Pang, M. Hehn, S. Alebrand, M. Cinchetti, G. Malinowski, Y. Fainman, M. Aeschlimann, and E. E. Fullerton, Nature Mater. **13**, 286 (2014).
85 I. Radu, K. Vahaplar, C. Stamm, T. Kachel, N. Pontius, H. A. Dürr, T. A. Ostler, J. Barker, R. F. L. Evans, R. W. Chantrell, A. Tsukamoto, A. Itoh, A. Kirilyuk, Th. Rasing, and A. V. Kimel, Nature **472**, 205 (2011).
86 D. Hinzke, U. Atxitia, K. Carva, P. Nieves, O. Chubykalo-Fesenko, P. M. Oppeneer, and U. Nowak, Phys. Rev. B **92**, 054412 (2015).
87 A. V. Kimel, A. Kirilyuk, P. A. Usachev, R. V. Pisarev, A. M. Balbashov and Th. Rasing, Nature 435, 655-657 (2005)
88 M. Berritta, R. Mondal, K. Carva, P. M. Oppeneer, arXiv:1604.01188.

[89] T.A. Ostler et al. Nature Commu. **3**, 666 (2011).
[90] C-H. Lambert, S. Mangin, B. S. D. Ch. S. Varaprasad, Y. K. Takahashi, M. Hehn, M. Cinchetti, G. Malinowski, K. Hono, Y. Fainman, M. Aeschlimann, and E. E. Fullerton. Science **345**, 1337 (2014).
[91] D. Hinzke, U. Nowak, P. Openeneer, unpublished; Y.K. Takahashi, R. Medapalli, S. Kasai, J. Wang, K. Ishioka, S.H. Wee, O. Hellwig, K. Hono, E.E. Fullerton, arXiv:1604.03488.
[92] M. S. El Hadri, P. Pirro, C.-H. Lambert, N. Bergeard, S. Petit-Watelot, M. Hehn, G. Malinowski, F. Montaigne, Y. Quessab, R. Medapalli, E.E. Fullerton, and S. Mangin, Appl. Phys. Lett. **108**, 092405 (2016).
[93] P. Elliott, K. Krieger, J. K. Dewhurst, S.Sharma, E. K. U. Gross, New J. Phys. 18, 013014 (2016).
[94] J. Simoni, M. Stamenova, S. Sanvito arXiv:1604.06262.
[95] P. Elliott, T. Müller, J. K. Dewhurst, S. Sharma, E. K. U. Gross, arXiv:1603.05603.
[96] R. J. Elliott, Phys. Rev. **96**, 266 (1954).
[97] M. Fähnle, J. Seib, and C. Illg, Phys. Rev. B **82**, 144405 (2010).
[98] K. Carva, M. Battiato, and P. Oppeneer, Phys. Rev. Lett. **107**, 207201 (2011).
[99] Y. Yafet, Solid State Phys. **14**, 1 (1963).
[100] R. N. Edmonds, M. R. Harrison, and P. P. Edwards. Annu. Rep. Prog. Chem., Sect. C: Phys. **82**, 265 (1985).
[101] F. Beuneu, and P. Monod, Phys. Rev. B **18**, 2422 (1978); P. Monod and F. Beuneu, ibid. **19**, 911 (1979).
[102] B. M. Khabibullin, and É.G. Kharakhash'yan, Sov. Phys. Usp. **16**, 806 (1974).
[103] V. Kamberský, Phys. Rev. B **76**, 134416 (2007).
[104] M. Münzenberg and J.S. Moodera, Phys. Rev. B **70**, 060402(R) (2004).
[105] M. B. Stearns, J. Magn. Magn. Mater. **5**, 167 (1977).
[106] J. Walowski, G. Müller, M. Djordjevic, M. Münzenberg, M. Kläui, C. A. F. Vaz, and J. A. C. Bland, Phys. Rev. Lett. **100**, 237401 (2008).
[107] I. Radu, G. Woltersdorf, M. Kiessling, A. Melnikov, U. Bovensiepen, J.-U. Thiele, and C. H. Back, Phys. Rev. Lett. **102**, 117201 ( 2009).
[108] G. Müller, J. Walowski, M. Djordjevic, G.-X. Miao, A. Gupta, A. V. Ramos, K. Gehrke, V. Moshnyaga, K. Samwer, J. Schmalhorst, A. Thomas, A. Hütten, G. Reiss, J. S. Moodera, M. Münzenberg, Nature Mater. **8**, 56 (2009).
[109] D. Steil, O. Schmitt, R. Fetzer, T. Kubota, H. Naganuma, M. Oogane, Y. Ando, A. K. Suszka, O. Idigoras, G. Wolf, B. Hillebrands, A. Berger, M. Aeschlimann and M. Cinchetti, New. J. Phys. **16**, 063068 (2014).
[110] D. Steil, et al. Phys. Rev. Lett. **105**, 217202 (2010).
[111] D. Steil, PhD thesis, University of Kaiserslautern 2012.
[112] A. Mann, J. Walowski, M. Münzenberg, S. Maat, M. J. Carey, J. R. Childress, C. Mewes, D. Ebke, V. Drewello, G. Reiss, and A. Thomas, Phys. Rev. X **2**, 041008 (2012).
[113] J. F. Cooke, J. W. Lynn, and H. L. Davis Phys. Rev. B **21**, 4118 (1980); J. A. Blackman, T. Morgan, and J. F. Cooke, Phys. Rev. Lett. **55**, 2814 (1985); D. M. Paul, P-W. Mitchell, H. A. Mook, and U. Steigenberger, Phys. Rev. B **38**, 580 (1988).
[114] A. T. Costa, Jr., R. B. Muniz, and D. L. Mills, Phys. Rev. B **69**, 064413 (2004).
[115] J. Mendil, P. Nieves, O. Chubykalo-Fesenko, J. Walowski, T. Santos, S. Pisana and M. Münzenberg, Sci. Rep. **4**, 3980 (2014).
[116] E. Turgut, C. La-o-vorakiat, J. M. Shaw, P. Grychtol, H. T. Nembach, D. Rudolf, R. Adam, M. Aeschlimann, C. M. Schneider, T. J. Silva, M. M. Murnane, H. C. Kapteyn, S. Mathias, Phys. Rev. Lett. **110**, 197201 (2013).
[117] B. Y. Mueller, A. Baral, S. Vollmar, M. Cinchetti, M. Aeschlimann, H. C. Schneider, and B. Rethfeld. Phys. Rev. Lett. **111**, 167204 (2013)
[118] A. Manchon, L. Xu Q. Li, and S. Zhang. Phys. Rev. B **85**, 064408 (2012).
[119] J. Fidler, and T. Schrefl, J. Phys. D **33**, R135 (2000).
[120] U. Nowak, O. N. Mryasov, R. Wieser, K. Guslienko, and R. W. Chantrell, Phys. Rev. B **72**, 172410 (2005).
[121] N. Kazantseva, U. Nowak, R. W. Chantrell, J. Hohlfeld, and A. Rebei, Europhys. Lett. **81**, (2007).

[122] U. Atxitia, O. Chubykalo-Fesenko, J. Walowski, A. Mann and M. Münzenberg, Phys. Rev. B **81**, 174401 (2010).
[123] U. Atxitia, O. Chubykalo-Fesenko, N. Kazantseva, D. Hinzke, U. Nowak, and R. W. Chantrell, Appl. Phys. Lett. **91**, 232507 (2007).
[124] M. Djordjevic and M. Münzenberg, Phys. Rev. B **75**, 012404 (2007).
[125] N. Kazantseva, D. Hinzke, U. Nowak, R. W. Chantrell, and O. Chubykalo-Fesenko, Phys. Stat. Sol. **244**, 4389 (2007).
[126] N. Kazantseva, D. Hinzke, U. Nowak, R. W. Chantrell, U. Atxitia, and O. Chubykalo-Fesenko, Phys. Rev. B **77**, 184428 (2008).
[127] D. A. Garanin, and O. Chubykalo-Fesenko, Phys. Rev. B **70**, 212409 (2004); D. A. Garanin, Phys. Rev. B **55**, 3050 (1997).
[128] W. F. Brown, Phys. Rev. **130**, 1677 (1963); O. Chubykalo, J. D. Hannay, M. A. Wongsam, R. W. Chantrell, and J. M. Gonzalez, Phys. Rev. B **65**, 184428 (2002).
[129] D. A. Garanin, Physica A **172**, 470, (1991).
[130] D. A. Garanin, Phys. Rev. B **55**, 3050 (1997).
[131] U. Atxitia, P. Nieves and O. Chubykalo-Fesenko, Phys. Rev. B **86**, 104414 (2012).
[132] O. Chubykalo-Fesenko, U. Nowak, R. W. Chantrell, and D. Garanin, Phys. Rev. B **74**, 094436 (2006).
[133] R. F. L. Evans, D. Hinzke, U. Atxitia, U. Nowak, R. W. Chantrell, and O. Chubykalo-Fesenko, Phys. Rev. B **85**, 014433 (2012).
[134] O. Chubykalo-Fesenko, U. Nowak, R. W. Chantrell, and D. Garanin, Phys. Rev. B **74**, 094436 (2006).
[135] U. Atxitia, O. Chubykalo-Fesenko, R. W. Chantrell, U. Nowak, and A. Rebei, Phys. Rev. Lett. **102**, 057203 (2009)
[136] C. Vogler, C. Abert, F. Bruckner, and D. Suess, Phys. Rev. B **90**, 214431 (2014).
[137] B. Koopmans, G. Malinowski, F. Dalla Longa, D. Steiauf, M. Fähnle, T. Roth, M. Cinchetti, and M. Aeschlimann, Nature Mater. **9**, 259 (2010).
[138] U. Atxitia and O. Chubykalo-Fesenko, Phys. Rev. B **84**, 144414 (2011).
[139] J. Kimling, J. Kimling, R. B. Wilson, B. Hebler, M. Albrecht, and D. G. Cahill, Phys. Rev. B **90**, 224408 (2014).
[140] A. I. Lobad, R. D. Averitt, C. Kwon, and J. Taylor, Appl. Phys. Lett. **77**, 4025 (2000).
[141] T. Hickel, B. Grabowski, F. Körmann, and J. Neugebauer, J. Phys.: Condens. Matter **24**, 053202 (2012).
[142] J. H. Mentink, J. Hellsvik, D. V. Afanasiev, B. A. Ivanov, A. Kirilyuk, A. V. Kimel, O. Eriksson, M. I. Katsnelson, and T. Rasing. Phys. Rev. Lett. **108**, 057202 (2012).
[143] L. Xu and S. Zhang. J. Appl. Phys. **113**,163911 (2013).
[144] C. Abert, M. Ruggeri, F. Bruckner, C. Vogler, G. Hrkac, D. Praetorius and D. Suess, Sci. Rep. **5**, 14855 (2015).
[145] M. Teichmann, B. Frietsch, K. Döbrich, R. Carley, and M. Weinelt. Phys. Rev. B **91**, 014425 (2015).
[146] T. Roth, A. J. Schellekens, S. Alebrand, O. Schmitt, D. Steil, B. Koopmans, M. Cinchetti, M. Aeschlimann
Phys. Rev. X **2**, 021006 (2012).
[147] U. Bovensiepen, Nature Phys. **5**, 461 (2009).
[148] G. P. Zhang, W. Hubner, G. Lefkidis, Y. Bai, and T.F. George, Nature Phys. **5**, 499 (2009).
[149] C. Vicario, C. Ruchert, F. Ardana-Lamas, P. M. Derlet, B. Tudu, J. Luning and C. P. Hauri, Nature Photon. **7**, 720 (2013); M. Shalaby, C. Vicario, C. P. Hauri, arXiv:1506.05397 (2015).
[150] J.-Y. Bigot, M. Vomir, and E. Beaurepaire, Nature Phys. **5**, 515 (2009).
[151] A. V. Kimel et al., Phys. Rev. B **63**, 235201 (2001).
[152] M. Pohl et al., Phys. Rev. B **88**, 195112 (2013).
[153] V. V. Kruglyak et al., Phys. Rev. B **71**, 233104 (2005); R. Wilks, R.J. Hicken, J. Phys. : Cond. Mat. **16**, 4607, (2004).
[154] F. Boschini, M. Mansurova, G. Mussler, J. Kampmeier, D. Grützmacher, L. Braun, F. Katmis, J. S.

Moodera, C. Dallera, E. Carpene, C. Franz, M. Czerner, C. Heiliger, T. Kampfrath, M. Münzenberg, Sci. Rep. **5**, 15304 (2015).
[155] P. B. Corkum and F. Krausz. Nature Phys. **3**, 381 (2007).